\documentclass[twocolumn,aps,prx,superscriptaddress]{revtex4-1}
\usepackage{graphicx}
\usepackage{color}
\usepackage{setspace}
\usepackage{xr}
\usepackage{chngcntr}
\usepackage{here}
\begin{document}

\title{Common mechanism of thermodynamic and mechanical origin for ageing and crystallisation of glasses}

\date{\today}

\author{Taiki Yanagishima}
\affiliation{Department of Fundamental Engineering, Institute of Industrial Science, The University of Tokyo, Komaba 4-6-1, Meguro-ku, Tokyo, 153-8505, Japan}
\author{John Russo}
\affiliation{Department of Fundamental Engineering, Institute of Industrial Science, The University of Tokyo, Komaba 4-6-1, Meguro-ku, Tokyo, 153-8505, Japan}
\affiliation{School of Mathematics, University of Bristol, Bristol BS8 1TW, United Kingdom }
\author{Hajime Tanaka}
\email{tanaka@iis.u-tokyo.ac.jp}
\affiliation{Department of Fundamental Engineering, Institute of Industrial Science, The University of Tokyo, Komaba 4-6-1, Meguro-ku, Tokyo, 153-8505, Japan}

\begin{abstract}
{\bf
The glassy state is known to undergo slow structural relaxation, where the system progressively explores lower free-energy minima which are either amorphous (ageing) or crystalline (devitrification). Recently, there is growing interest in the unusual intermittent collective displacements of a large number of particles known as ``avalanches''. However, their structural origin and dynamics are yet to be fully addressed. Here, we study hard-sphere glasses which either crystallise or age depending on the degree of size polydispersity, and show that a small number of particles are thermodynamically driven to rearrange in regions of low density and bond orientational order. This causes a transient loss of mechanical equilibrium which facilitates a large cascade of motion. Combined with previously identified phenomenology, we have a complete kinetic pathway for structural change which is common to both ageing and crystallisation. Furthermore, this suggests that transient force balance is what distinguishes glasses from supercooled liquids.
}
\end{abstract}

\maketitle
\section*{Introduction}
Ageing and devitrification are very slow dynamical processes taking place in glasses. 
Ageing leads to a gradual change in the physical properties of glasses \cite{struik1978physical,kob1999computer}; a prominent example is a change in sample size, potentially very harmful for applications to high precision devices. Devitrification, the transformation of a glass to a crystal, has also been a subject of intensive study for many years~\cite{kelton2010nucleation}. For material scientists, the tendency for a supercooled system to devitrify during ageing or on heating can be detrimental to the stability of organic \cite{craig1999relevance,Powell2014} and metallic glasses \cite{greer1995metallic}, silicates \cite{Marshall1961}, macromolecules \cite{Sear2007,Doye2007} and aqueous systems \cite{angell2002liquid}. The latter is particularly relevant to cryogenics, where ice formation can have adverse effects on biological samples \cite{mazur1984freezing}. For many industrial products, including pharmaceuticals \cite{craig1999relevance}, devitrification during storage is also a very serious problem. Despite the technological importance of these phenomena, their fundamental mechanisms remain elusive, because of their slow dynamics and apparently stochastic nature.

To access the microscopic mechanisms behind ageing and devitrification, hard-sphere (colloidal) glasses provide an ideal system due to the experimental accessibility to particle-level information \cite{weeks2000three,gasser2001real,Simeonova2006,lynch2008dynamics,yunker2009irreversible} and the simple nature of the hard-core interaction. Different studies have considered ageing and devitrification separately in hard-sphere glasses. In the case of ageing, an increase in structural order over time was found~\cite{Kawasaki2014}. For hard spheres (both monodisperse and weakly polydisperse), structural order is characterised not only by density but also by bond orientational order (BOO), a measure of angular order between neighbouring particles. This ordering originates from a thermodynamic driving force to lower the free energy of the system. Note that higher BOO means larger vibrational (or correlational) entropy, i.e. lower free energy for hard spheres (see, e.g., ref. \cite{Tanaka2012}). Regarding devitrification, on the other hand, Pusey {\it et al.} \cite{Pusey2009} showed that crystallisation can take place beyond the glass transition point in over-compressed hard sphere glasses. This may apply more generally, as jammed or glassy metastable states are also seen to undergo ordering.

On deeper overcompression, Sanz {\it et al.} \cite{Sanz2014} found that crystallisation occurs through discrete collective events, where groups of particles suddenly undergo large displacements, accompanied by an equally sudden increase in the proportion of crystalline particles in the system: these events were termed `avalanches' for their intermittent, collective nature. They firstly found that the randomisation of particle velocities in their molecular dynamics (MD) simulation averted the incidence of the events, and secondly, that polydispersity suppressed the growth of crystallinity while maintaining the same intermittent dynamics. Thus, they concluded that avalanches `mediate' crystallisation, or more precisely, that `chance' collective motion in the system seemed to trigger the displacements that led to crystallisation. At the same time, they noted that there was a spatial heterogeneity to which such events are linked, and suggested `soft spots' \cite{Widmer-Cooper2008} as a potential candidate for what might distinguish them from other regions. So while ageing is proposed to be thermodynamically driven~\cite{Kawasaki2014}, it is argued that devitrification and avalanche events are of a strongly stochastic nature with an unidentified structural signature for initiation; crystallisation is their by-product. Thus, the connection between ageing and devitrification, if any, remains elusive: despite being the two principal dynamic phenomena taking place in glasses, their physical mechanisms also remain unknown. For example, what makes the kinetics of the phenomena in glasses special remains unclear, since the relation of ageing and devitrification to corresponding processes in the supercooled liquid state, i.e. structural relaxation and crystallisation respectively, are yet to be ascertained. 

Here, we address these fundamental problems by studying structural evolution during avalanches in both ageing and devitrifying hard-sphere systems. We show that ageing and devitrification are both characterized by the same kinetic processes, and that the special feature of dynamics in glasses originates from a temporal mechanical balance characteristic of solids, which has not been clearly recognized so far. We distinguish the very first particles to take in avalanche events (AIs) and go on to show that they are characterised by a structural signature, with a lower local density and low bond orientational order. We then show that this motion leads to a transient loss in the force balance of the system, and it is this that gives rise to the cascade of particle motion which we perceive as an `avalanche'. Finally, we find that this motion triggers displacements throughout the system in the vicinity of pre-ordered regions. We believe this to correspond to the avalanche devitrification identified by Sanz {\it et al.}, though we go on to show that this happens in ageing systems as well.

\section*{Results}

\begin{figure*}%
\includegraphics[width=14cm]{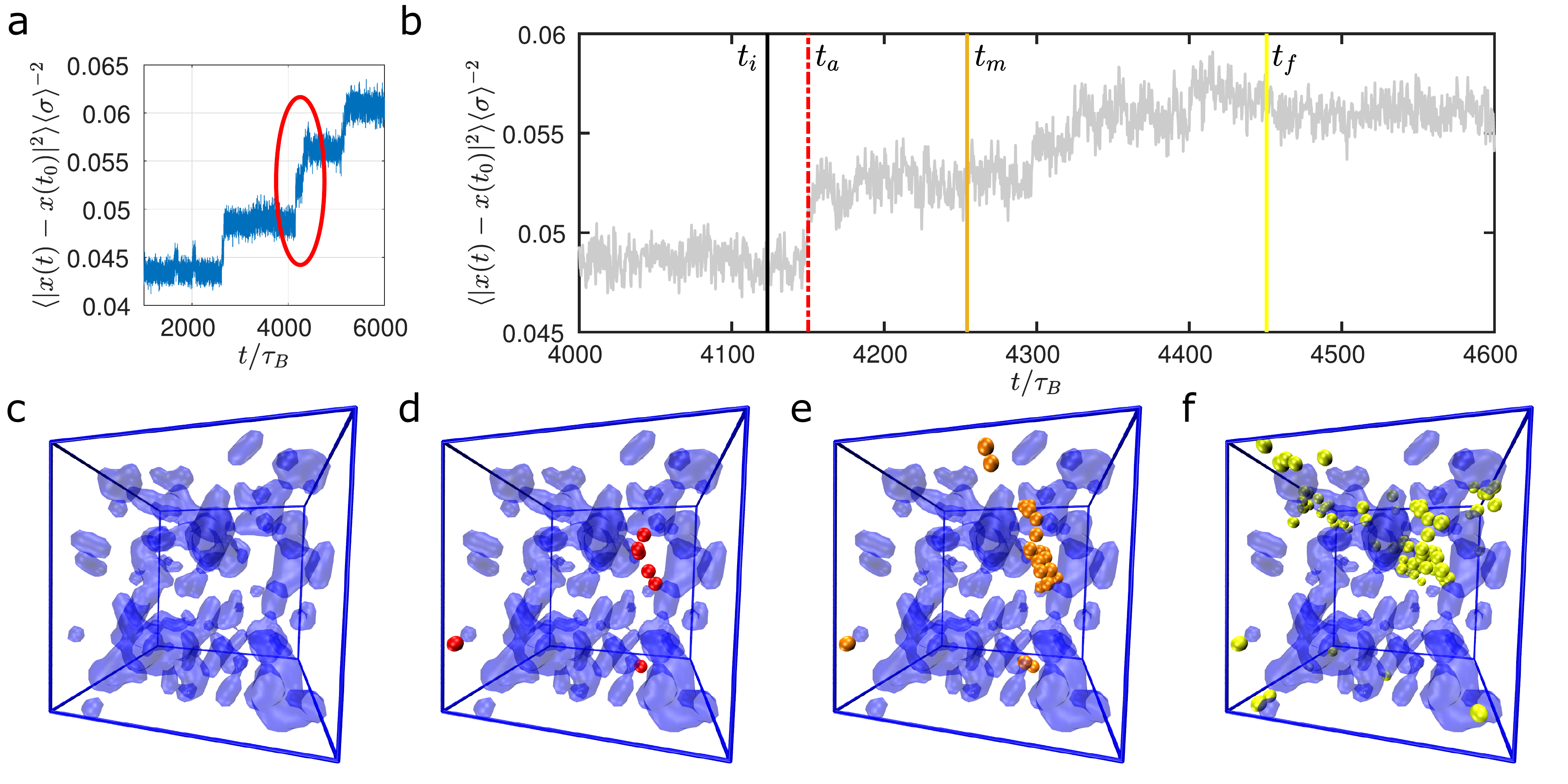}%
\caption{{\bf Avalanche over time seen in an ageing system at $\Phi=0.65$.}
{\bf a,} MSD over all particles from a time $t_0$. We can see distinct avalanche events that happen intermittently. The circled event is the same as that shown in {\bf c-f}. {\bf b,} MSD over all particles from a time $t_0$, focused on the avalanche event circled in {\bf a}. Times $t_i$, $t_a$, $t_m$ and $t_f$ correspond to snapshots {\bf c} to {\bf f} respectively. {\bf c-f,} Progression of the circled avalanche event in panel {\bf a}. Blue shaded regions are regions where $Q_6 > 0.25$. We only display particles which have been displaced more than $\sigma_l/3$ since snapshot {\bf c} in each panel. We can see the number of these particles increase progressively. Particles displayed at panel {\bf d} can be regarded as AIs, whereas particles appearing later on are APs. All other particles are not displayed. Despite the step-like appearance of the MSD in {\bf b}, it is reasonable to assume that this is one collective event, as we can see in panels {\bf c-f}.} %
\label{fig:MSD}%
\end{figure*}

\noindent
{\bf Progression of an avalanche.} 
We simulate supercooled monodisperse and weakly polydisperse hard-sphere-like systems in the glassy regime to study devitrification and ageing, respectively. Details regarding the system and the simulation are given in the section of Brownian Dynamics simulation in Methods. For the monodisperse case, we see a clear increase in the crystallinity, but not for the polydisperse case, as shown in Supplementary Fig.~1. We proceed to study how an avalanche event is triggered in a glass state, how this triggering event spreads over the system, or induces an avalanche, and how structural ordering is finally enhanced by the avalanche. We firstly observe the incidence and progression of avalanches, looking at the mean squared displacement of particles in the system from some arbitrary time $t_0$. An example is shown in Fig.~\ref{fig:MSD}a for an ageing system. We can see that the sudden collective displacement of particles takes place intermittently. We proceed to look at the circled event in more detail in Fig.~\ref{fig:MSD}b. Firstly, we identify four different times during the event: a time just preceding the event, $t_i$; the time at which the first particles are displaced and the MSD suddenly rises, $t_a$; an intermediate time $t_m$, purely for illustrative purposes; the time at which the MSD finally reaches its next plateau, $t_f$. Snapshots of the system at each of these times are given in Fig.~\ref{fig:MSD}c-f, with blue shaded regions corresponding to regions with high BOO parameter $Q_6 > 0.25$ (see the section of Structural analysis in Methods for the definition of BOO) at time $t_i$. Only particles which are displaced more than $\sigma_l/3$ ($\sigma_l$ being the size parameter of particle $l$ (see the section of Analysis of avalanche events in Methods) between $t_i$ and the time of the snapshot are shown.

Starting from a quiescent, metastable state in panel c, we see that a small, localised cluster of these particles appear in panel d. When the number of nearby particles displaced more than the $\sigma_l/3$ threshold becomes greater than 5 for the first time after time $t_i$, we call these particles AIs (avalanche initiators). These will be discussed at length below. Note that they appear in disordered regions with low local crystalline BOO (low $Q_6$). This is followed by particles in the vicinity being subsequently displaced, as shown in panel e, before the system reaches the next plateau at panel f. The avalanche particles identified at the end of the event are called APs (avalanche particles), and, with few exceptions, are significantly more numerous than AIs.

\begin{figure*}%
\includegraphics[width=12cm]{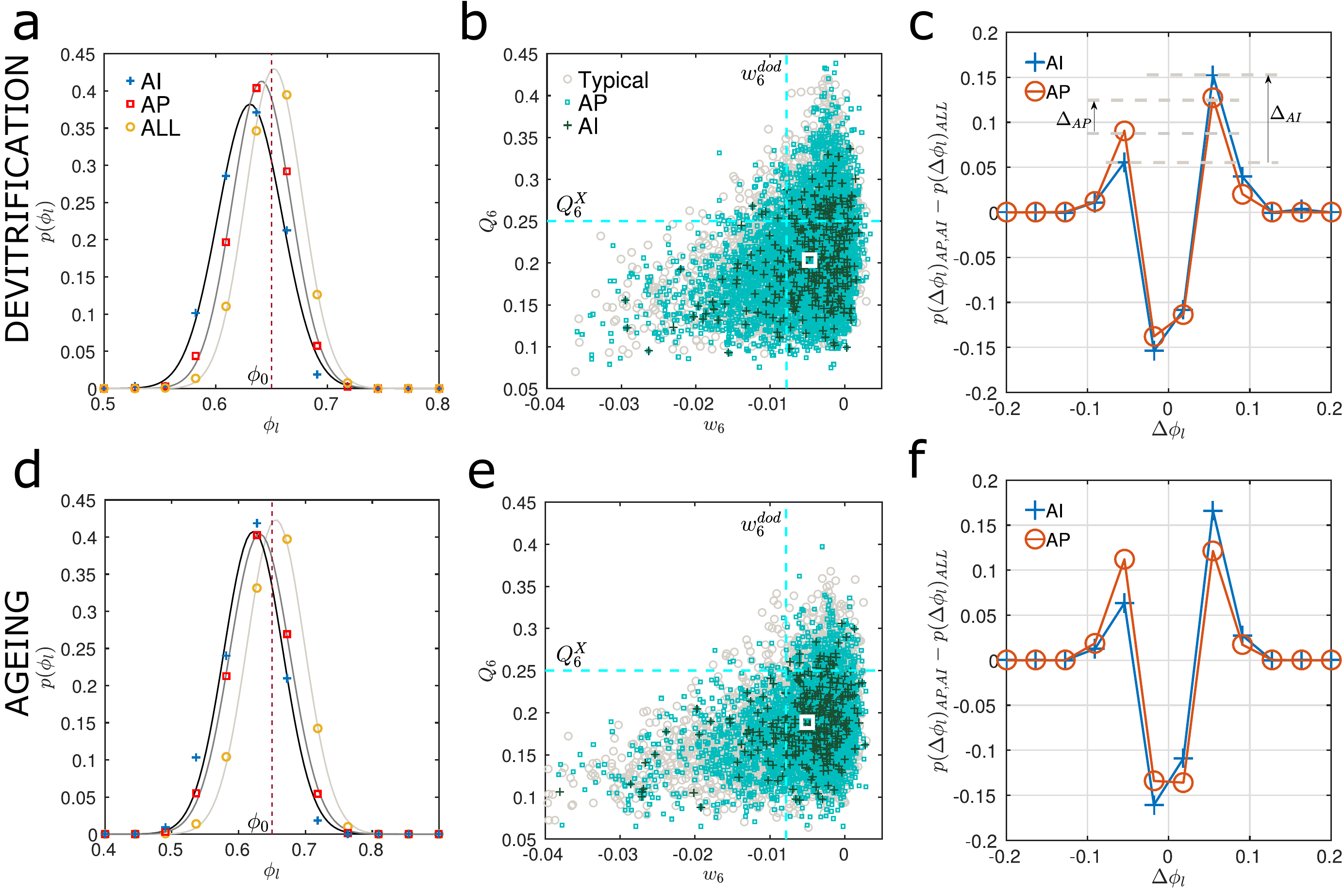}%
\caption{{\bf Characterisation of avalanche events.}
Averages are given over 50 events in both devitrifying and ageing events. {\bf a, d,} Distribution of local volume fraction $\phi_l$ before avalanche events. Both show a peak shift to lower $\phi_l$ for AIs, less so for APs. {\bf b, e,} $w_6$ and $Q_6$ distributions for AIs and APs (defined in text)  overlaid on a typical distribution for all particles. The AIs especially are concentrated at lower BOO parameter values (both $w_6$ and $Q_6$).  Dashed line $Q_6^X$ and $w_6^{dod}$ show the threshold values separating particles with highly crystal-like order ($Q_6 \geq Q_6^X$) and those with high dodecagonal or icosahedral-like order ($w_6 \leq w_6^{dod}$) \cite{Leocmach2012a}. Note that crystals have $Q_6$ larger than 0.4 (see the section of Structural analysis in Methods for details). The white thick open square represents the average values of $w_6$ and $Q_6$ for AIs. On average, thus, AIs possess neither significant crystalline order nor locally favoured structures. This indicates that AIs are particles localized in disordered regions. {\bf c, f,} Difference between the probabilities of $\Delta\phi_l$ for AI or AP subsets and the distribution for the system as a whole. There is a bias towards larger change for both AIs and APs, but there is a skew to positive charge for AIs which is not seen for APs. For both devitrification and ageing, APs almost equally gain or lose density, but AIs have a strong tendency to increase local density.}%
\label{fig:Stats}%
\end{figure*}

\noindent
{\bf Structural origin and evolution of AIs and APs.} 
Note that we now have 3 populations of particles, AIs, APs and the whole set. These can be analysed separately with a large enough sample of events. With lists of AIs and APs over all the events, we go on to look at the typical local structure of AIs and APs at $t_i$. We start with local volume density $\phi_l$,  defined as $\frac{1}{6}\pi d_l^3/V^l_{vor}$, where $d_l$ is the effective particle diameter (see the section of Brownian Dynamics simulation in Methods) and $V_{vor}^l$ is the local volume of particle $l$ as found from a radical Voronoi tesselation. A distribution of $\phi_l$ values for AIs, APs and all particles at $t_i$, is shown in Fig.~\ref{fig:Stats}a and d for devitrifying and ageing systems respectively. It is clear that AIs are found in regions with a markedly lower density than the global average. APs exhibit this to a much lesser extent.

A lower density can arise from either larger local Voronoi volumes or simply smaller particle size. In the monodisperse (devitrifying) case, only the former is possible, whereas both are possible for the polydisperse case. We look at the distribution of local Voronoi volumes for the polydisperse (ageing) case, only to find that the volume for AIs is in fact skewed towards lower values, i.e., towards {\it higher} volume fraction, albeit only slightly. This is shown in Supplementary Fig.~2a, where the distributions of local Voronoi volumes are given for AIs, APs and all particles. The lower local volume fraction in the polydisperse case must thus come primarily from AIs having smaller particle size, as confirmed in the particle size distributions given in Supplementary Fig.~2b.

Similar intermittent dynamics have been reported for ageing in glasses in simulations and experiments  \cite{ElMasri2010,Vollmayr-Lee2006,Evenson2015,Fan2015}. These works identified avalanche events as barrier-crossing events in a complex energy landscape under the influence of internal stresses. In this energy-landscape picture, AIs and APs can be related to the activation stage (from the initial states to the nearby saddle states) and the relaxation stage (from the saddle states to the final states), respectively. Egami and his coworkers~\cite{Fan2015} also found cascade events spreading over the whole system in unstable glasses formed by an instant quench, which may be caused by large-amplitude density inhomogeneities frozen in by such a quench. A link between such nonlocal events and self-organized criticality was also suggested \cite{Vollmayr-Lee2006,Fan2015}, but it is beyond the scope of our paper. Here we focus on localized events in a rather stable glass, which is crucial for identifying structural characteristics that are independent of the box size. This allows us to reveal a specific mechanism for event initiation, propagation and consequence.

Going on to look at BOO parameters $Q_6$ and $w_6$, we superimpose the values of these parameters for AIs and APs over all trajectories for devitrifying and aging cases over a typical distribution over the whole system at some arbitrary time point (see Fig.~\ref{fig:Stats}b and e). $Q_6$ is associated with crystalline BOO in hard sphere systems, while a lower (more negative) $w_6$ is associated with dodecahedral and icosahedral BOO \cite{Leocmach2012a}. It is clear that AIs can be found concentrated in regions where neither $Q_6$ nor $-w_6$ are notably large, i.e. in disordered regions. The average value over the AIs is noted as a white square on each, both below the thresholds associated with crystals and locally favoured structures. The distribution of APs, on the other hand, are indistinguishable from the typical distribution.

\begin{figure}%
\begin{center}
\includegraphics[width=0.9\columnwidth]{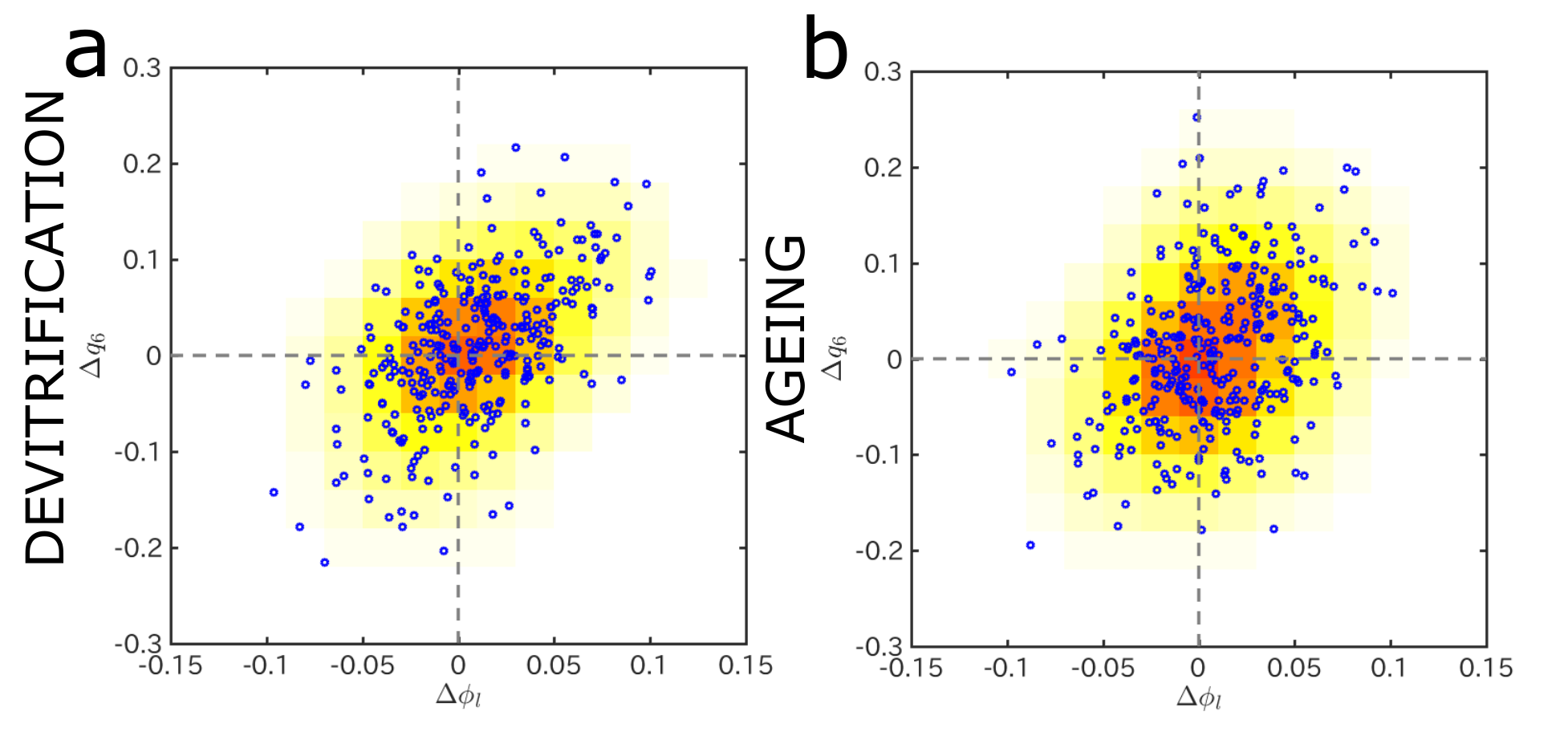}%
\end{center}
\caption{{\bf Correlation between $\Delta\phi_l$ and $\Delta{q}_6$.}
Distribution of $\Delta\phi_l$ and $\Delta{q}_6$ for AIs for {\bf a,} Devitrification, {\bf  b,} Ageing. There is a clear positive correlation between translational and orientational order at this depth of overcompression ($\Phi=0.65$).
}%
\label{fig:Q6rho}%
\end{figure}

From these two findings, we can conclude that avalanches are more likely to initiate from a low degree of structural order and a low local Voronoi density (see Fig.~\ref{fig:Q6rho} on the correlation between the structural order and local density). This region may be regarded as a kind of ``defect'' in a glass. 

We look further, and see whether certain changes in translational order or BOO are more likely for AIs and APs compared to the population as a whole. This can be shown by calculating $p(\Delta{O})_{subset} - p(\Delta{O})_{whole}$, where $O$ is the order parameter of interest. Looking at the devitrifying case first in Fig.~\ref{fig:Stats}c, we find several things. Firstly, it is clear that both AIs and APs, with their large displacements, generally entail a greater change in the local volume fraction $\phi_l$ than average, giving the curves their distinctive `cursive V' shape. AIs and APs are less likely to have a local volume fraction change $|\Delta \phi_l|<0.04$ than average, and more likely to have a $\phi_l$ change greater than 0.04, with a maximum around $|\Delta \phi_l| \sim 0.06$. This implies that particles with large amplitude displacements are most likely accompanied by a local volume change of magnitude $\sim$0.06 under dense packing. Secondly, we note that the AIs are very clearly skewed towards an increase in local volume fraction $\phi_l$. What is particularly interesting is that this feature is {\it not} observed for APs, i.e.  they are equally likely to find a positive and negative local volume change. We will show later that this is a direct consequence of the fact that the motion of APs is of mechanical (not thermodynamic) origin. Looking to the ageing case in Fig.~\ref{fig:Stats}f, we once again see this asymmetry for AIs and APs. The statistical robustness of these results is discussed in the section of Event statics in Methods.

It is worth noting that local structural order and local density are not entirely independent of each other at such a high packing density \cite{Tanaka2012}. We are clearly in a regime where translational ordering and bond orientational ordering are closely related to each other, particularly for hard-sphere-like systems, where the local symmetry is selected by packing. Instead of looking at the structure at $t_i$, we now look at the change in local volume fraction $\phi_l$ and BOO over the time interval $t_i < t < t_f$ for the three subsets, starting with AIs. In order to sharpen the sensitivity of the BOO parameter to localised change, we choose $q_6$ instead of its coarse-grained counterpart $Q_6$. As seen in Fig.~\ref{fig:Q6rho}a and b, we see a positive correlation between an increase in $q_6$ and $\phi_l$. Note that the colour map simply reflects the number of particles in the plot for particular $\Delta\phi$ and $\Delta{q_6}$ ranges. It should be noted that the trend is much stronger for the devitrifying case.

As seen in Fig.~\ref{fig:Stats}, AIs originate from regions of lower local volume fraction (or, density) and low BOO. Thus, the increase in $\phi_l$ is not so much a densification beyond the system average, but rather a relaxation of a density inhomogeneity present in the heterogeneous, over-compressed structure. The causal relationship between them is clearly highlighted by the difference in the degree of asymmetry seen between AIs and APs. Since AIs have a structural origin, their states before and after are distinct. There is thus a clear structural distinction between AIs and APs: the asymmetry in structural evolution of AIs indicates the presence of a local, thermodynamic driving force for the initiation of avalanche events. On the other hand, it would appear that APs are rather a kinetic by-product of the AIs (discussion continued below). Here, it is worth noting that it is known \cite{Conrad2005} that there is no discernible difference between the distributions of Voronoi volumes and mobility for a supercooled liquid: it is our ability to identify AIs and APs that uncovers this structural distinction.

\begin{figure}%
\begin{center}
\includegraphics[width=0.9\columnwidth]{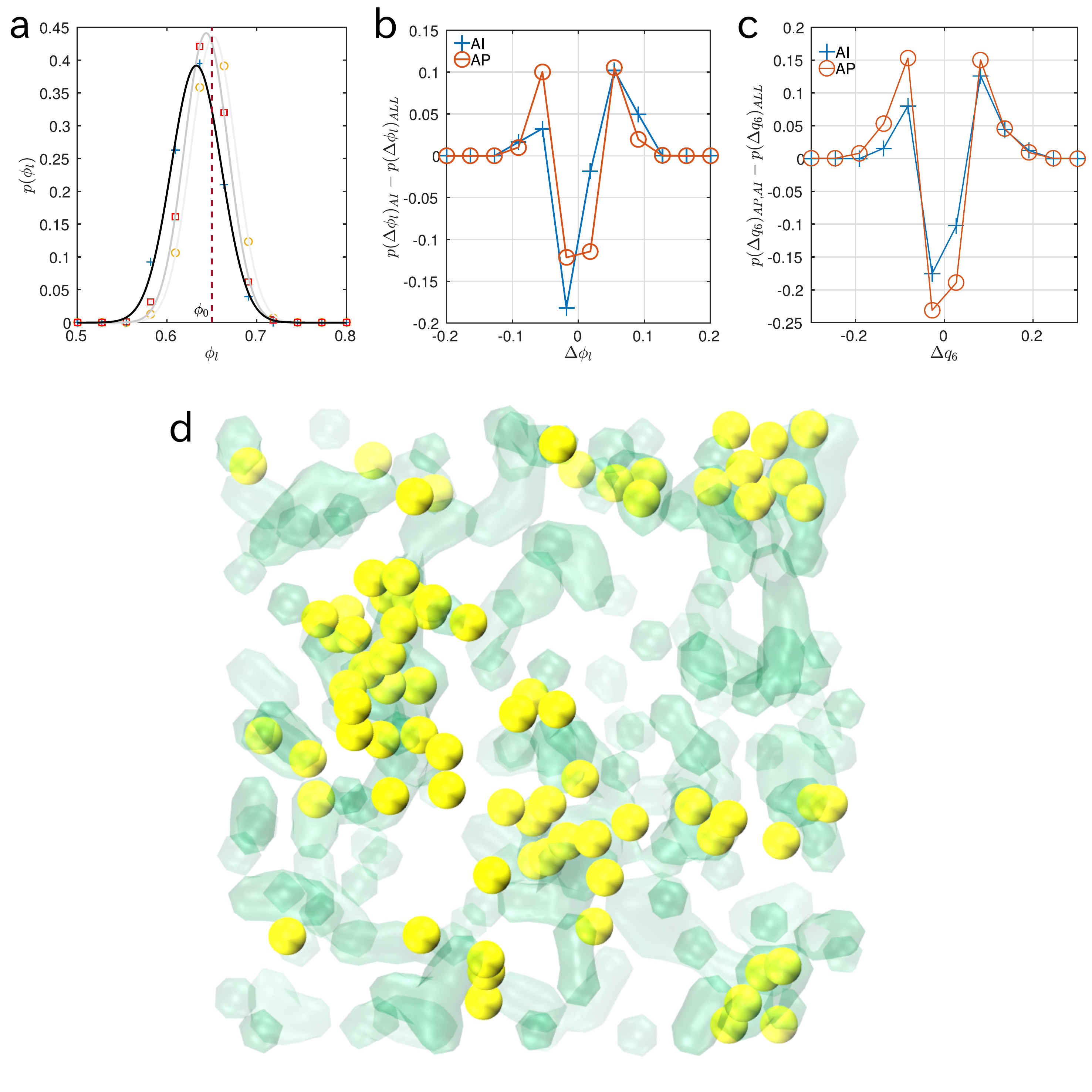}%
\end{center}
\caption{{\bf Iso-configurational ensemble.}
Analysis of 38 events found in 100 iso-configurational trajectories for the monodisperse case ($\Phi=0.65$). The initial state is taken from one of the MSD plateaux in the original 50 independent runs. {\bf a,} Local Voronoi density distribution for AIs, APs and all particles, and differences between all-particle probability distributions and AI, AP distributions for changes in ({\bf b}) local density $\phi_l$ and ({\bf c}) bond-orientational order parameter $q_6$. Data is given for the monodisperse set. Note that all prominent features of the independent runs are replicated, including a lower initial local density and marked asymmetry in structural change for AIs, and not for APs. {\bf d,} A map of all unique AIs (yellow) from the detected events and regions (transparent green) which fulfil all of the following conditions: $Q_6 < 0.25$, $w_6 > -0.01$, $\phi_l < 0.62$.
}%
\label{fig:iso}%
\end{figure}

These conclusions are further confirmed for the monodisperse case by using an iso-configurational ensemble. We produced an iso-configurational ensemble (approx. 2000$\tau_B$ length), initiated from a configuration taken from one of the many MSD plateaux in the monodisperse case. 100 trajectories were found, and 38 events were recorded. Further to reproducing the spatial heterogeneity identified in \cite{Sanz2014}, we apply the same treatment as above to produce Fig.~\ref{fig:iso}a-c: the trends are exactly the same as what was found for 50 independent trajectories. As seen in Fig.~\ref{fig:iso}d, there is good agreement between the unique AIs (yellow) and regions which obey all of the following criteria: $Q_6 < 0.25$, $w_6 > -0.01$, $\phi_l < 0.62$, where the AI $\phi_l$ probability distribution becomes larger than the distribution over the whole system. Low density and low BOO thus provide us with the regions where AIs are more likely to appear, as they are characterized by lower activation energies for particle motion, where the caging effect is locally weaker. As in any stochastic process, the study of regions of low density and low BOO order only provides a probability map of where AIs will occur. The actual location cannot be predicted in a deterministic way, as the underlying motion is Brownian, without memory effects (inertia).

\begin{figure}%
\includegraphics[width=0.9\columnwidth]{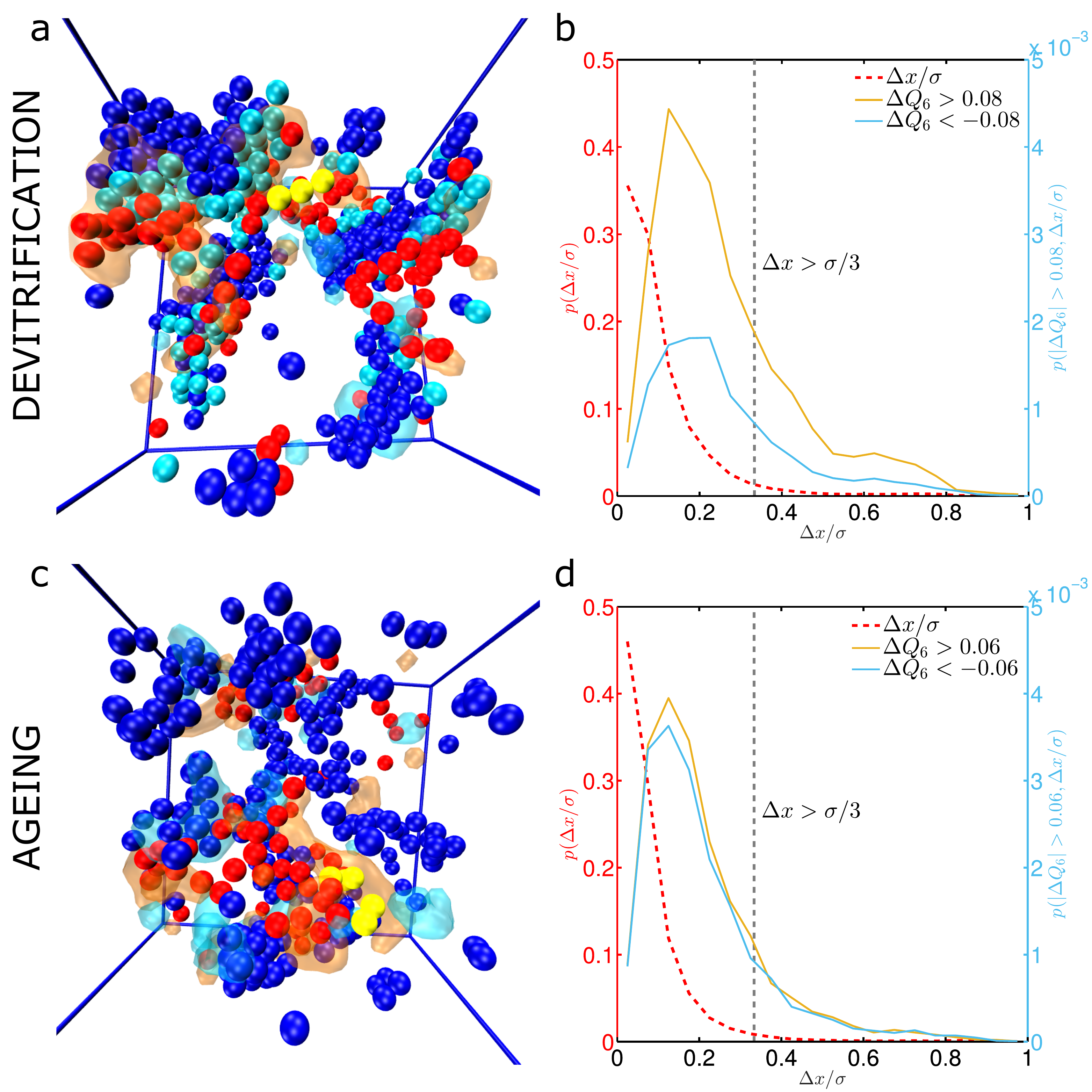}%
\caption{{\bf Displacement and structural change.}
{\bf a,} Spatial distribution of crystalline (blue), newly crystalline (light blue), AIs (yellow) and APs (red) during an avalanche event in the monodisperse case at $\Phi=0.65$. Regions which have experienced a large change in BOO, ${\Delta}Q_6 > 0.08$ and ${\Delta}Q_6 < -0.08$, are marked with a transparent orange and cyan region, respectively. {\bf b,} Fraction of particles with displacement $\Delta x/\sigma$ for particles experiencing a large change in BOO parameter $Q_6$ during avalanches for monodisperse and polydisperse cases respectively.  The red dashed curve is for all particles (left axis), whereas the orange and cyan curves are for particles experiencing a large positive and negative change in $Q_6$ respectively (right axis).  The threshold is the same as in a. {\bf c,} The spatial distribution of MRCO regions (blue), AIs (yellow) and APs (red) during an avalanche event in the polydisperse case at $\Phi=0.65$. Regions which have experienced a large change in BOO, ${\Delta}Q_6 > 0.06$  and ${\Delta}Q_6 < -0.06$, are marked as a transparent orange and cyan region, respectively. {\bf d,} Same as b for the ageing case. The threshold is the same as in c. Note that the peaks in both b and d occur at displacements smaller than the avalanche threshold, $\Delta{x} > \sigma / 3$, for particles experiencing a large change in $Q_6$. 
}%
\label{fig:DSDQ}%
\end{figure}

\noindent
{\bf Function of APs in structural evolution.} 
Does this mean that APs do not have any role to play in crystallisation or ageing? If the only significant structural evolution in the system was derived from the AIs, the total change would be minuscule, as AIs associated with an event make up only around 0.2 to 0.3\% of the system (approx. 10 particles in 4000). Consider the events shown in Fig.~\ref{fig:DSDQ}, where panels a and c are events from devitrifying and ageing trajectories, respectively. Yellow particles are AIs and red particles are APs. In panel a, dark blue particles are particles which are crystalline at $t_i$, and light blue particles are those that become crystalline by $t_f$. In panel c, blue particles are those with $Q_6 > 0.25$. Orange and blue transparent clouds are regions with a pronounced increase or decrease in $Q_6$, respectively, thresholded at 0.08 in panel a and 0.06 in panel c. It is clear that the incidence of APs causes a significant amount of structural change in different parts of the system, particularly in the vicinity of pre-ordered clusters: importantly, regions where the {\it largest} BOO development is seen are {\it not} necessarily clustered around the AIs or APs. 

This last point is worth considering carefully. Though there is some overlap, the particles which in fact experience the largest change are those which move {\it less} than the avalanche threshold, but more than average. Fig.~\ref{fig:DSDQ}b and d show the fraction of particles with a particular displacement size for particles undergoing a large change in $Q_6$, for monodisperse and polydisperse cases, respectively, using the same thresholds for $Q_6$ changes as in panel a and c. The distribution of displacements for particles which undergo large structural change are peaked below the avalanche threshold, $\Delta{x} > \sigma_l/ 3$. This is not simply a reflection of the overall distribution of displacements in the system (see the red dashed curve), which has no peak and monotonically decreases with an increase in the displacement. Also note that there is a clear bias towards positive changes (orange curve) in $Q_6$ compared to negative changes (blue curve) for the devitrifying case.

In the above we have shown that APs are not the particles which crystallise, as also reported in Ref.~\cite{Sanz2014}. 
In the final stage of the avalanche identified in our work, APs cause movements which are smaller than the avalanche threshold and yet induce significant BOO development. 
This raises the fundamental question of what is the physical mechanism behind the avalanche ‘mediated’ 
devitrification \cite{Sanz2014} in monodisperse systems, or why such small particle displacements (less than the particle size) can facilitate crystallization. 
The natural answer to this question arises from a crystallization pathway where the first step is development of spatial coherency in bond orientational order and the fact that small displacements are enough for this enhancement \cite{Russo2012}. On noting that enhancement of BOO spatial coherency cannot induce translational ordering in systems with a high enough polydispersity \cite{Tanaka2012}, we can now also claim the same mechanism for ageing systems.

In supercooled liquid and glassy states with weak frustration against crystallisation, like the one studied here, the system always seeks a way to lower the free energy even at a local scale by using every possible kinetic path. In a glassy state, a system tries to increase structural order locally, which results in density inhomogeneities accumulated in the disordered region overcoming the energy barrier to motion (AIs), causing a cascade of particle motion (APs).

\noindent
{\bf Intermittent dynamics and mechanical equilibrium.} 
Having ascertained the structural origins of avalanche initiation, an important question arises: what are the conditions underlying the intermittency of the dynamics? In an ordinary supercooled liquid, particle motion proceeds in a continuous manner. It is only in a highly supercooled or glassy state that we see intermittent, avalanche-like particle motion. We argue that the latter is a characteristic of a glassy state, where {\it mechanical equilibrium} is satisfied both locally and globally, at least transiently. Suggestions have been made \cite{goldstein1969viscous,dyre1999solidity} and evidence given \cite{lemaitre2014structural,gelin2016anomalous} that mechanical balance is attained in a glassy state or in the inherent structure, resulting in the emergence of a long-range elastic correlation like in jammed granular matter \cite{henkes2009statistical}. Before an avalanche, a system is in a mechanically stable state even with thermal noise: the local ordering events which trigger 
the motion of AIs lead to the breakdown of this transient mechanical equilibrium, or the loss of force balance, until the system attains a new mechanically stable state. 

\begin{figure}[tp]%
\includegraphics[width=0.9\columnwidth]{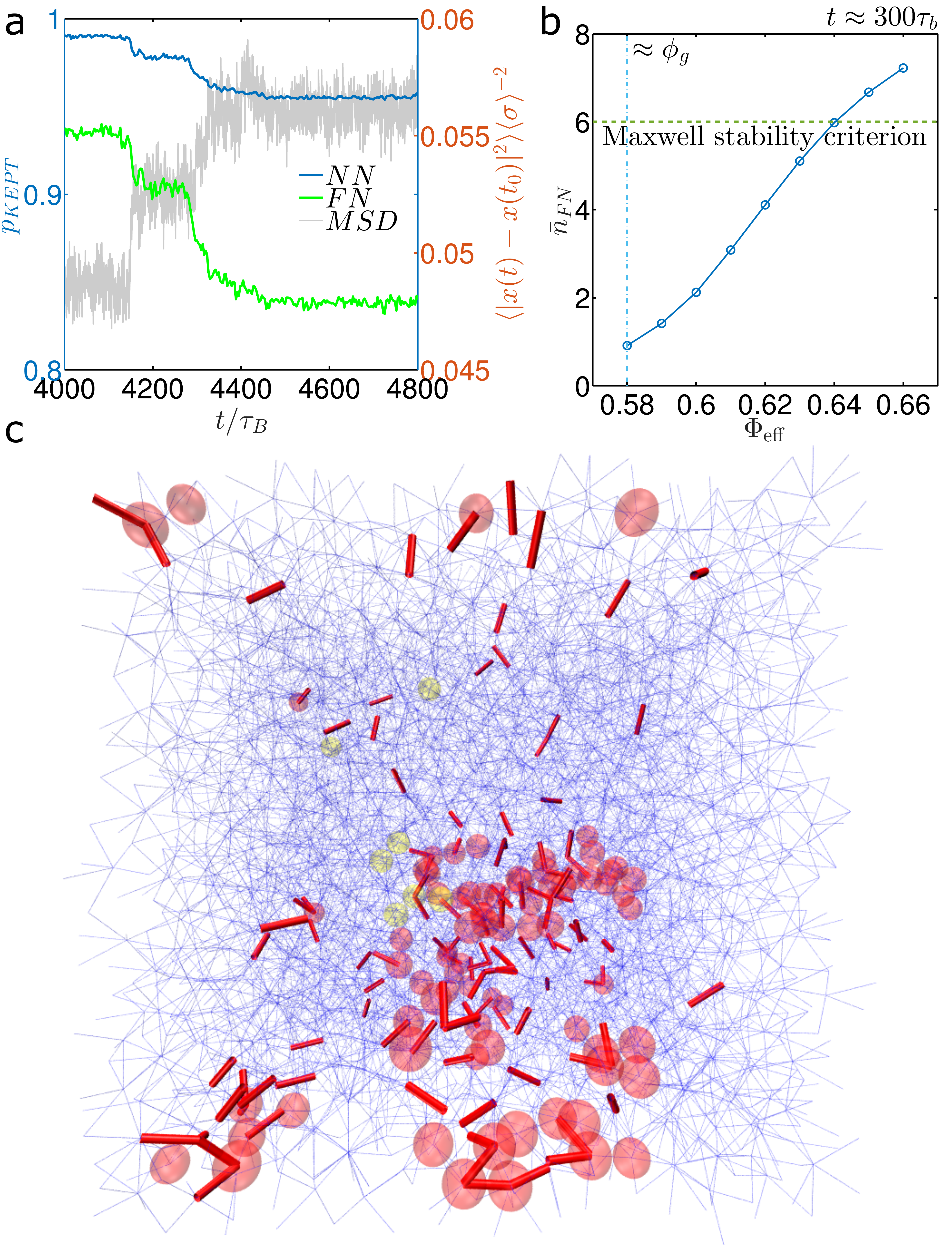}%
\caption{{\bf Breakage of force chains.}
{\bf a,} The average proportion of neighbours kept by particles (left axis) compared to its neighbours at some initial time preceding an avalanche event in a polydisperse trajectory ($\Phi=0.65$). Results are shown for both nearest neighbours (NNs: blue curve) and force neighbours (FNs: green curve). The mean squared displacement of the system is shown in the background by a grey curve (right axis): the re-arrangement of the system force network coincides with the avalanche. {\bf b,} The average number of force neighbours, $\bar{n}_{FN}$, observed after $300\tau_B$ at different volume fractions. {\bf c,} Re-arrangement of the force network during the avalanche event shown in {\bf a}. Yellow and red particles are AIs and APs respectively, as in Fig.~\ref{fig:DSDQ}a and c. The stable force network that is not involved in the re-arrangement is shown by the blue network, whereas newly formed force bonds resulting from a re-arrangement since a time preceding the avalanche are 
shown as thick red bonds. We can see some correlation between APs and AIs and a significant local re-arrangement of FNs. This is a snapshot after an avalanche event; please see the Supplementary Movie to see the correlation over the course of the whole event.
}%
\label{fig:Forces}%
\end{figure}

Evidence for this in our system is given in Fig.~\ref{fig:Forces}, where we look at how particles change their neighbours over time. We can define two types of neighbour, simply using proximity (nearest neighbours, NNs) or those with which a repulsive force is acting (force neighbours, FNs). The average number of neighbours kept compared to some initial time is shown in Fig.~\ref{fig:Forces}a, both for FNs and NNs, superimposed on the progression of an avalanche event in a polydisperse trajectory. We can immediately see that the system is in a state of mechanical equilibrium before, with an extended plateau over which particles keep its neighbours, and maintains the force chain structure of the system. It is only when the avalanche initiates that this structure changes. The existence of a clear plateau afterwards shows that mechanical equilibrium has been restored, with a different connectivity between particles as before.

The best evidence for the role of mechanical stability is spatial correlation between AIs/APs and new FN connections. This is shown in Fig.~\ref{fig:Forces}c, where new force connections formed over the course of an avalanche have been shown as bold red bonds, superimposed on transparent blue bonds representing the original force network before the event. The AIs (yellow) and APs (red) for the event are also shown. The abrupt formation of these new FNs is also shown dynamically in the Supplementary Movie.

It is worth noting that the average number of FNs surrounding AIs and APs is around 6, the Maxwell rigidity criterion \cite{Maxwell1865} in 3 dimensions. Distributions for the number of FNs for AIs, APs and all particles for both monodisperse and polydisperse cases are given in Supplementary Fig.~3. The role of this in the dynamics becomes clearer when different volume fractions are considered. Short trajectories are produced for volume fractions $\Phi$ ranging from 0.58 to 0.66 using the same number of particles $n$, and the average number of FNs recorded after $t\approx{300}\tau_B$. It can be seen that the average number of FNs, $\bar{n}_{FN}$, first passes this threshold when the volume fraction reaches $\Phi = 0.65$. This coincides with the volume fraction above which the avalanches can be observed, i.e. sudden jumps are seen in the MSD preceded and followed by clear plateaus (see Fig.~\ref{fig:MSDex}). Also note that the increase in $n_{FN}$ starts from the nominal glass transition volume fraction, $\phi_g \approx 0.58-0.59$, corroborating the role that mechanical contacts might have to play in glass formation.

Note that the frequency of these jumps also changes with volume fraction. Avalanche events take place in an intermittent manner when a system is able to overcome the barrier between one metastable basin and another with the help of thermal noise. Since the height of this barrier steeply increases with an increase in the degree of supercooling, the frequency of avalanches should decrease for deeper supercooling; we confirm this in our simulations, as shown in Supplementary Fig.~4, where avalanche incidence is increasingly suppressed at higher volume fractions. This trend is also seen qualitatively in \cite{Rosales-Pelaez2016} using a much sharper inter-particle potential.

\section*{Discussion}
We have now seen that avalanche initiation has a structural precursor, and that the propagation of APs and the intermittency of the dynamics is related to the temporary loss of mechanical stability of the system for both devitrifying and ageing systems. Avalanches further induce small-amplitude motion of particles (less than $\sigma_l/3$) in regions which already have a high structural order. It is however worth noting what makes the devitrifying and aging systems different. In the devitrifying systems considered here, the crystal growth is interface-limited, meaning that the barrier for crystallisation comes from the addition of particles to the crystalline surface. This process does not require large-scale translational motion~\cite{Zaccarelli2009a,Sanz2011,Russo2012}, and can thus occur under the small-amplitude motion set in place by AP particles. In most ageing systems (mixtures or highly polydisperse samples), crystal growth is instead diffusion-limited, thus requiring large-scale diffusional motion for the interface to grow, and it is thus not easily activated by avalanches motion. This is shown by the fact that particles undergoing large changes in BOO for the polydisperse case do not seem to reflect the clear bias to positive changes in $Q_6$ seen in the monodisperse case, as shown in Fig.~\ref{fig:DSDQ}b and d. This is also evident for AIs and APs in Supplementary Fig.~5a,b, where there is no clear bias to order development like in Fig.\ref{fig:Stats}c and f. These show that BOO development in ageing systems is not in the expected thermodynamic direction for every event. Of the 50 polydisperse runs, 32 cases experience an overall increase in $Q_6$, the others experience a decrease. There is however an overall positive trend: the mean change in $Q_6$ per event is $\Delta{Q}_6 = (2.24 \pm 1.94)\times{10}^{-4}$, where the error is an unbiased standard error in $\bar{\Delta{Q}_6}$. This corresponds to $\approx 87$\% confidence in a $Q_6$ increase. This is in contrast to a $\Delta{Q}_6 = (3.83 \pm 1.39)\times{10}^{-4}$ for the monodisperse case, with a 99.8\% confidence in a $Q_6$ increase. This bias to positive changes reflects the thermodynamic direction, which is hindered by the suppression of long-range diffusion in the polydisperse case.

As seen in Fig.~\ref{fig:DSDQ}c, regions where large changes in BOO occur are localised to pre-ordered regions (MRCO). Previous work on ageing in 3D hard sphere systems at lower volume fractions \cite{Kawasaki2014} has shown that there is a slow development in the spatial correlation length of BOO parameter $Q_6$. Such an increase is most easily achieved if re-structuring takes place near pre-ordered sites, explaining the large changes in $Q_6$ close to MRCO regions induced by nearby AP motion. This is analogous to the localisation of $Q_6$ growth to the MRCO in the devitrifying case, with an added stochasticity factor.

Finally, we give evidence for the fact that avalanche particles form strings, as briefly noted in \cite{Sanz2014}. Supplementary Fig.~6 shows the degree of overlap with the former position of neighbors at an arbitrarily chosen initial point over time, a decrease in which denotes string-like displacement. The exact definition is given in the section of String-like collective motion in Methods. The event is the same as the one used in Fig.\ref{fig:Forces}a; the coincidence with avalanche incidence is clear. It is particularly interesting that the collective, string-like motion seen in the Adam-Gibbs scenario for quasi-equilibrium supercooled liquids \cite{Donati1998} could be involved in the structural evolution of the glassy state of unstable glasses like the ones shown here. String-like motion may be the only possible way for particles to move in a highly packed state \cite{Donati1998}. This similarity in dynamics lends further credence to a thermodynamic picture of the glass transition and the intrinsic link between crystallisation and vitrification under the influence of frustration \cite{Tanaka2012}.

The emerging picture of an avalanche in glasses is thus the following. Avalanches occur between mechanically stable states, where the force network between particles does not change with time. They are initiated by small clusters of particles (AIs, avalanche initiators) that have lower than average local density, within environments with a low crystalline bond-orientational order, a lack of locally favoured structures (icosahedra), and where the number of constraints is close to the threshold for mechanical stability. In polydisperse systems, AIs are more often small particles. Due to the loss of mechanical stability, the displacement of the AIs is propagated through the system in a cascade-like fashion; hese particles are called APs (avalanche particles) and, on average, do not directly experience distinct structural evolution, i.e. locally, they will almost equally likely decrease or increase the amount of crystalline order in the system, although slightly biased towards the increase. This makes them unique from AIs. However, the APs induce significant change in their surroundings which move less, but cause a development of $Q_6$ in the vicinity of nearby highly ordered clusters. This triggering of ordering is analogous to the crystallisation of supercooled liquids under some external perturbation. In short, we find that the ageing and devitrification of glasses proceed via a common pathway involving the interplay of thermodynamics and mechanics: thermodynamically-initiated mechanical facilitation of particle motion (i.e., avalanche) and the enhancement of structural order by the cascade of small-scale displacements induced by this avalanche event.

This work highlights a kinetic pathway via which strongly supercooled glassy systems can age or devitrify, highlighting the particular role that inhomogeneities in density and structural order have to play in their stability - a stronger frustration towards this kind of relaxation would make a glassy state more stable against ageing or devitrification. This is a useful guiding principle for the stabilisation of the amorphous state in various materials. For a multi-component glass, for example, local demixing is required for devitrification. The mechanism of devitrification in such a mixture remains for future investigation.


\vspace{5mm}

\noindent
{\bf Acknowledgements}
\noindent
This study was partly supported by Grants-in-Aid for Specially Promoted Research and for Young Scientists (KAKENHI Grants No. 25000002 and No. 15K17734) from the Japan Society for the Promotion of Science (JSPS). TY is a Research Fellow (PD) of the Japan Society for the Promotion of Science and also gratefully acknowledges the Postdoctoral Fellowship by Special Examination (2015-6) of the Institute of Industrial Science, University of Tokyo.

\noindent
{\bf Author Contributions} 
\noindent
H.T. conceived and supervised the project, T.Y. performed the numerical simulations and analysis. J.R. wrote the BOO calculation packages. All authors discussed the results and contributed to the writing of the manuscript.

\noindent
{\bf Author Information} 
\noindent
Correspondence and requests for materials should be addressed to H.T.

\noindent
{\bf Competing Financial Interests} 
\noindent
The authors do not declare any competing financial interests.


\nonumber


\vspace{10mm}

\clearpage
\section*{Methods}

\noindent
{\bf Brownian Dynamics simulation.}
We simulated supercooled monodisperse and weakly polydisperse glasses in the glassy regime using Brownian Dynamics (BD) simulation for particles interacting through a Weeks-Chandler Andersen (WCA) potential \cite{Weeks1971} in a three dimensional box with periodic boundary conditions. We employed the monodisperse system as an example of a devitrifying system and a weakly polydisperse system (6\% Gaussian) as an example of an ageing system \cite{Zargar2013}. The difference between them is clearly shown in Fig.~\ref{fig:Xt}, where the proportion of crystalline particles $p_X$ (see below) is shown over time. The devitrifying system experiences growth in crystallinity, whereas the ageing system does not.

Initial overcompressed states were attained using rapid compression, expanding the size of the particles from 1\% their intended size to their final dimension in less than a Brownian time $\tau_B$, including noise to allow frustration limited structures to relax. This is similar to the Lubachevsky-Stillinger algorithm \cite{Lubachevsky1990}. Although this volume fraction exceeds random close packing fraction ($\sim$64 \% and $\sim$66 \% for monodisperse and polydisperse cases, respectively \cite{Tanaka2012}), local structural ordering such as BOO allows a system to retain mobility. We note that the volume fraction is well above the glass transition volume fraction ($\Phi_{\rm g} \sim 0.58-0.59$) and structural relaxation in the system, including the time scale of avalanches, would be significantly longer than the Brownian time $\tau_B$.  As our primary interest is in persistent changes in structure as opposed to transient fluctuations, we average all positions and BOO 
parameters over 3$\tau_B$ intervals.

A Brownian Dynamics simulation code was developed in FORTRAN 95 using the Ermak-McCammon algorithm \cite{Ermak1978}, with no hydrodynamic interactions and no inertial term. Particles interact via a Weeks-Chandler-Andersen (WCA) potential, $U_{\rm WCA}(r) = 4{\epsilon}\{(\sigma/r)^{12} - (\sigma/r)^6 + 1/4\}$ for $r < 2^{\frac{1}{6}}\sigma$, where $\sigma$ is the diameter of the particle, and the temperature kept such that $k_{\rm B}T/\epsilon =$ 0.025 ($k_{\rm B}$ being the Boltzmann constant), as in ref.~\cite{Filion2011,Kawasaki2014}. 
Volume fractions are calculated using an effective diameter $d = 1.0953 \sigma$ where the WCA potential decays to $k_{\rm B}T$. This is in agreement with a mapping of the WCA potential phase diagram to the hard sphere one to within less than 0.5\% \cite{Ahmed2009}. The diffusivity $D$ is chosen to mimic experimental colloidal systems, assuming a 1 micron particle radius in aqueous viscous surroundings with a viscosity of 1 mPas.

For all simulations shown in the main text, the number of particles is $n = $ 4000 unless otherwise stated. 
Here we mention the dependence of the observed dynamics, particularly the MSD, on system size. We produced eight trajectories with a system size $n = 16000$, and compared their MSDs with eight randomly chosen trajectories from the $n = 4000$ set in the monodisperse case. This is given in Supplementary Fig.~7. It can be seen that events are smaller and less well-defined for larger system size. This is to be expected, since the properties of individual events should be independent of system size, i.e. since the MSD is normalized by particle number, and the events give smaller jumps in the MSD. Also, a larger system size makes it more likely for multiple events to be occurring at once, hence the lack of clear, flat plateaux, as events initiate before APs from another event have relaxed.

\noindent
{\bf Structural analysis.}
The metrics we employ include local density as characterised by a radical Voronoi construction, and the $q_6$ order parameter which has found wide acceptance as a suitable order parameter to detect crystalline order in hard spheres~\cite{Kawasaki2010,Russo2013}. 
Five-fold symmetric structures, such as icosahedral and dodecahedral structures, are targeted with the $w_6$ order parameter~\cite{Leocmach2012a}. BOO parameters such as $q_6$ and  $w_6$  are calculated as described in ref.~\cite{Steinhardt1983}. To reduce fluctuations in the distribution of $q_6$, a coarse-grained version, commonly written as $Q_6$, is also employed~\cite{Lechner2008}. All the definitions of the order parameters can be found in ref.~\cite{Russo2012}. Here we note that crystals have $Q_6$ larger than 0.4 \cite{Kawasaki2010,Kawasaki2010a}. The proportion of crystalline particles $p_X$ is defined as the proportion of particles in the system which have 7 or more `solid' bonds, i.e. bonds between nearest neighbors for which the bond coherence $d_{6ij} = \sum^{m=6}_{-6}{q}_{6m}(i){q}^\ast_{6m}(j)/|q_{6m}(i)||q_{6m}(j)|$ exceeds 0.7. Radical Voronoi constructions were carried out using the VORO++ package available at http://math.lbl.gov/voro++/ \cite{Rycroft2009}.

\noindent
{\bf Analysis of avalanche events.}
We analyse 3 populations of particles, AIs, APs and the whole set, with a large enough sample of events as follows. 
In 32 independent simulations of devitrifying and ageing systems, each approximately $7000{\tau_B}$ in length, we identified 50 avalanche events each for both cases. Care was taken to identify events which had a clear beginning and end, i.e. with a clear plateau in the mean-squared displacement at both ends. Each event was labelled manually with a $t_i$ and $t_f$, and APs were found by comparing the state of the system at these two times; AIs for each event were found by comparing particle positions for progressively later times from $t_i$ until the condition above was met. 

\noindent
{\bf Event statistics.}
We briefly comment on the statistics of this result. Histograms were generated by taking probability distributions over AIs, APs and ALL particles for individual events and then averaging these, avoiding bias of data to larger events amongst the 50 that were identified. The marker size (in both Fig. \ref{fig:Stats}c and f and Supplementary Fig.~5a and b) is an indication of the maximum contribution possible from each event. In total, 315 AIs and 6121 APs were counted for the monodisperse case, 320 AIs and 3718 APs for the polydisperse case. Given the relatively small number of AIs, the width of bins could not be made significantly wider than what is shown. In order to be sure that these trends are not due to some particularly favourable events mixed into the set, we focused on the two peaks in Fig.~2c and f, took a random selection of 25 events amongst the sets of 50, and measured the asymmetry between the AI and AP curves, marked $\Delta_{AI}$ and $\Delta_{AP}$ in the figure. This was repeated 100 times. In the monodisperse case, $\Delta_{AI} = 0.0976 \pm 0.0235$ and $\Delta_{AP} = 0.0384 \pm 0.0148$; in the polydisperse case, $\Delta_{AI} = 0.1044 \pm 0.025$ and $\Delta_{AP} = 0.0099 \pm 0.0108$. In both cases, $\Delta_{AI} > \Delta_{AP}$.

\noindent
{\bf String-like collective motion.}
Following the formalism of Donati {\it et al.} \cite{Donati1998}, we calculate the value of the collective motion parameter $\delta = \sigma_{ij}^{-1}\min \left({|X_i(t) - X_j(t_0)|,|X_j(t)-X_i(t_0)|}\right)$, averaged over nearest neighbours $i$ and $j$, and then over all particles in the system over time. $\sigma_{ij}$ is the sum of the radii of particles $i$ and $j$. The original work set a threshold for $\delta$ below which strings were considered to move collectively. Thus, we can interpret a decrease in the global average value of $\delta$ to be a collective displacement. As shown in Supplementary Fig.~6 (same polydisperse trajectory as in Fig.\ref{fig:Forces}a), the global value of $\delta$ decreases during the course of an avalanche, with well-defined metastable states before and after.s

\noindent
{\bf Data availability. }
The raw data analysed to derive the findings of this paper is available from the corresponding author upon reasonable request.

\clearpage
\centerline{\bf SUPPLEMENTARY FIGURES}

\setcounter{figure}{0}
\renewcommand{\figurename}{{\bf Supplementary Fig.}}
\renewcommand{\thefigure}{\arabic{figure}}

\begin{figure}[H]%
\centering
\includegraphics[width=6cm]{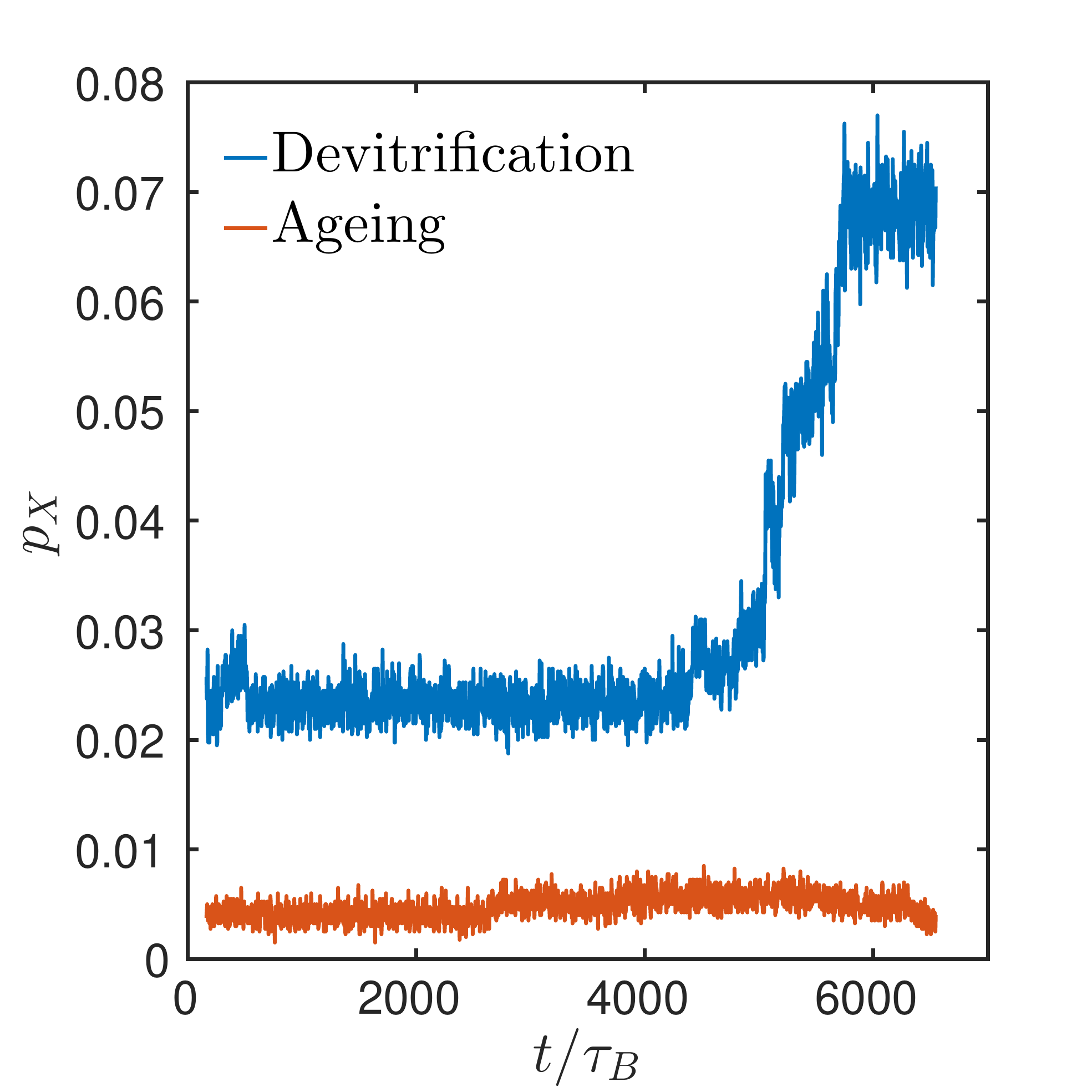}%
\caption{{\bf Crystallinity over time.}
The evolution of the proportion of crystalline particles $p_X$ over time for a randomly selected trajectory for the monodisperse, devitrifying case and the polydisperse, ageing case. Note that the crystallinity undergoes intermittent growth during devitrification but not during ageing.}
\label{fig:Xt}%
\end{figure}

\begin{figure}[H]%
\begin{center}
\includegraphics[width=8cm]{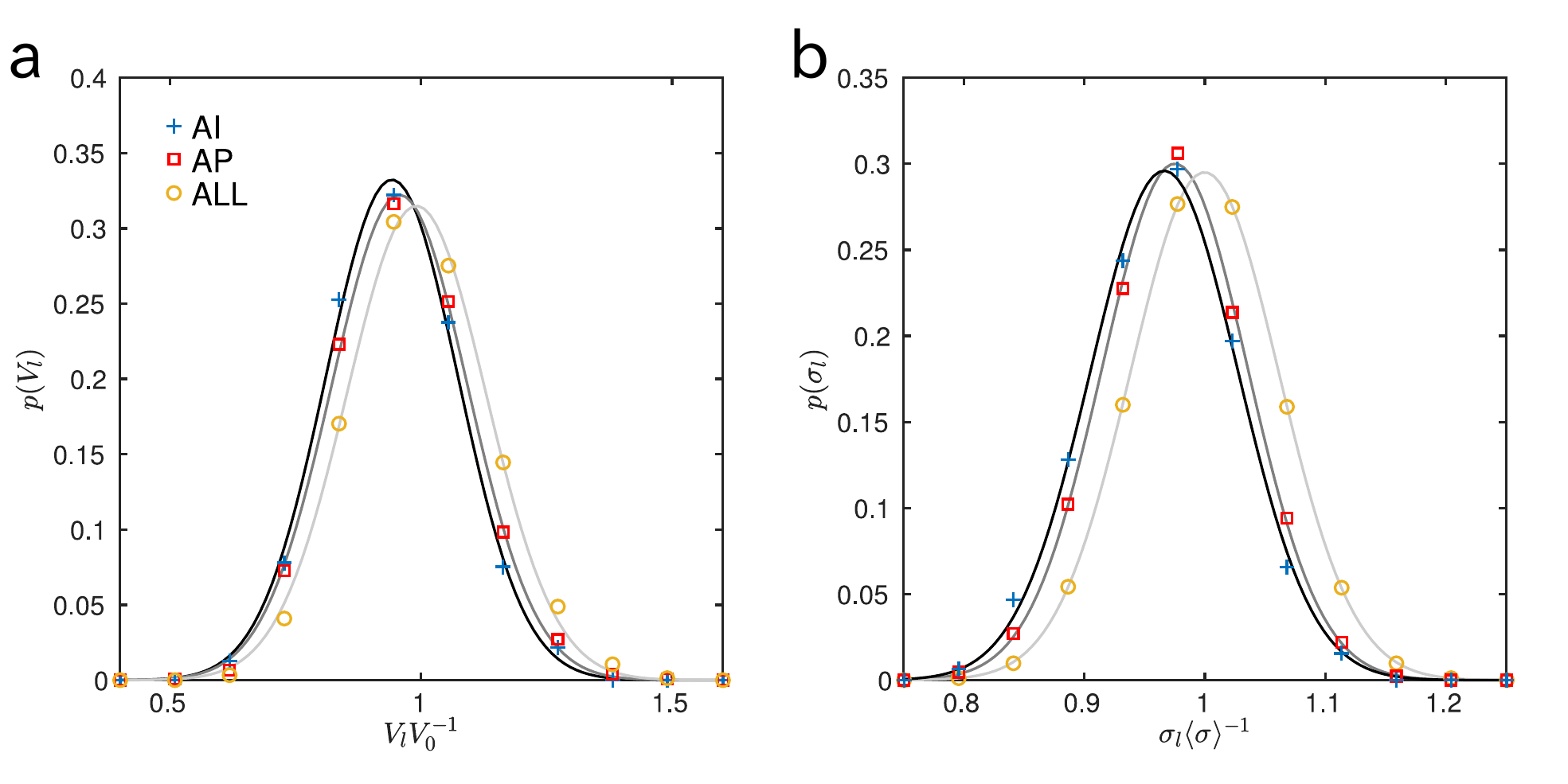}%
\end{center}
\caption{{\bf Voronoi volume or size?}
{\bf a,} Distribution of local Voronoi volume for AIs, APs and all particles in ageing polydisperse trajectories, averaged over 50 events. {\bf b,} Distribution of particle size for AIs, APs and all particles. Note that the resulting low local volume fraction for AIs and APs must be driven by the smaller size of particles, since the Voronoi volumes for AIs and APs are not clearly separated.
}%
\label{fig:VVS}%
\end{figure}

\begin{figure}[H]%
\begin{center}
\includegraphics[width=8cm]{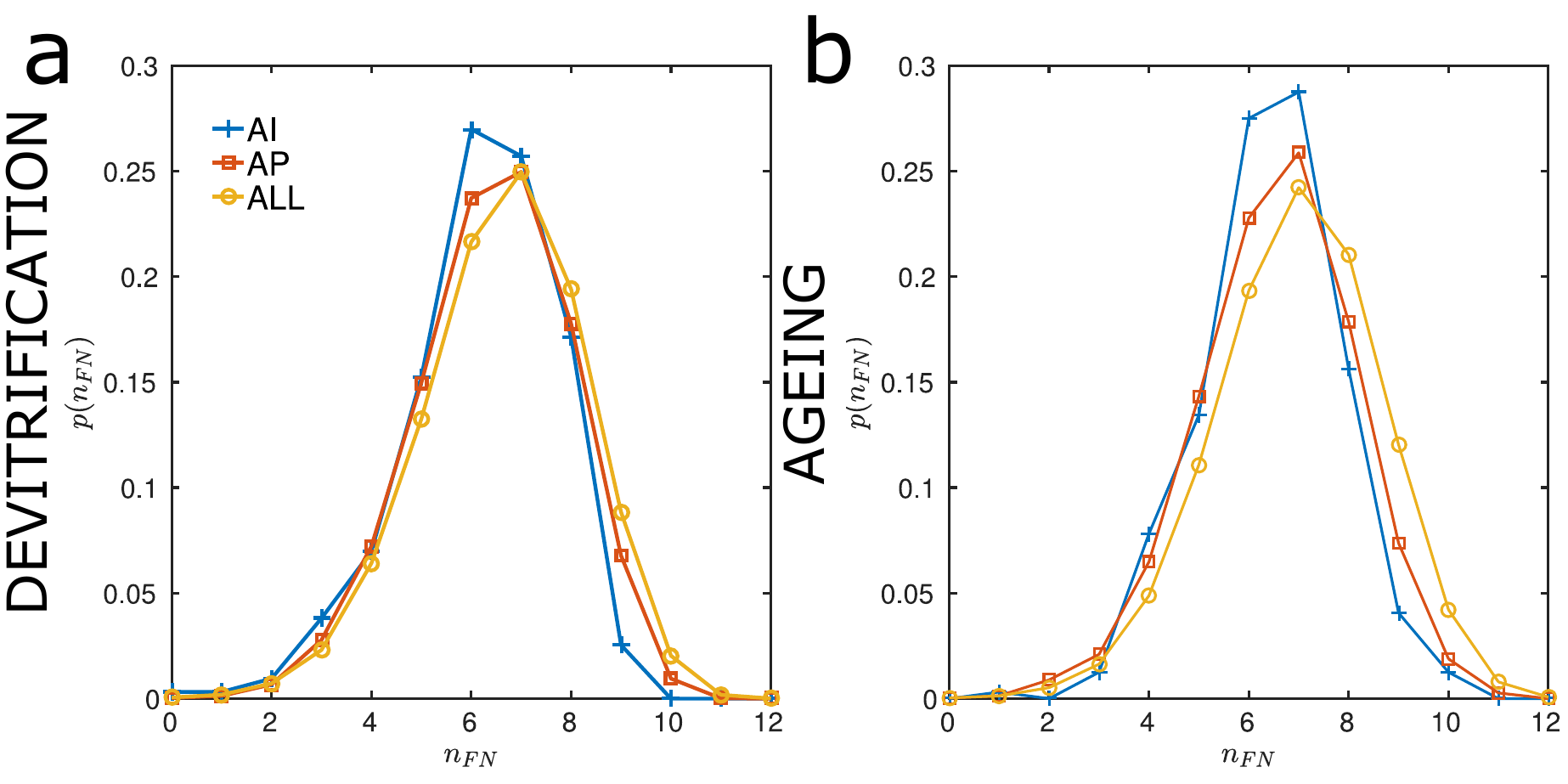}%
\end{center}
\caption{{\bf Force neighbours.}
Distribution of the number of force neighbours that particles have for AIs, APs and over all particles before avalanche events for the monodisperse ({\bf a}) and polydisperse ({\bf b}) cases. APs and AIs have a slightly smaller number of force neighbours in both cases, as expected from their lower Voronoi density.
}%
\label{fig:nFN}%
\end{figure}

\begin{figure}[h]
\begin{center}
\includegraphics[width=0.9\columnwidth]{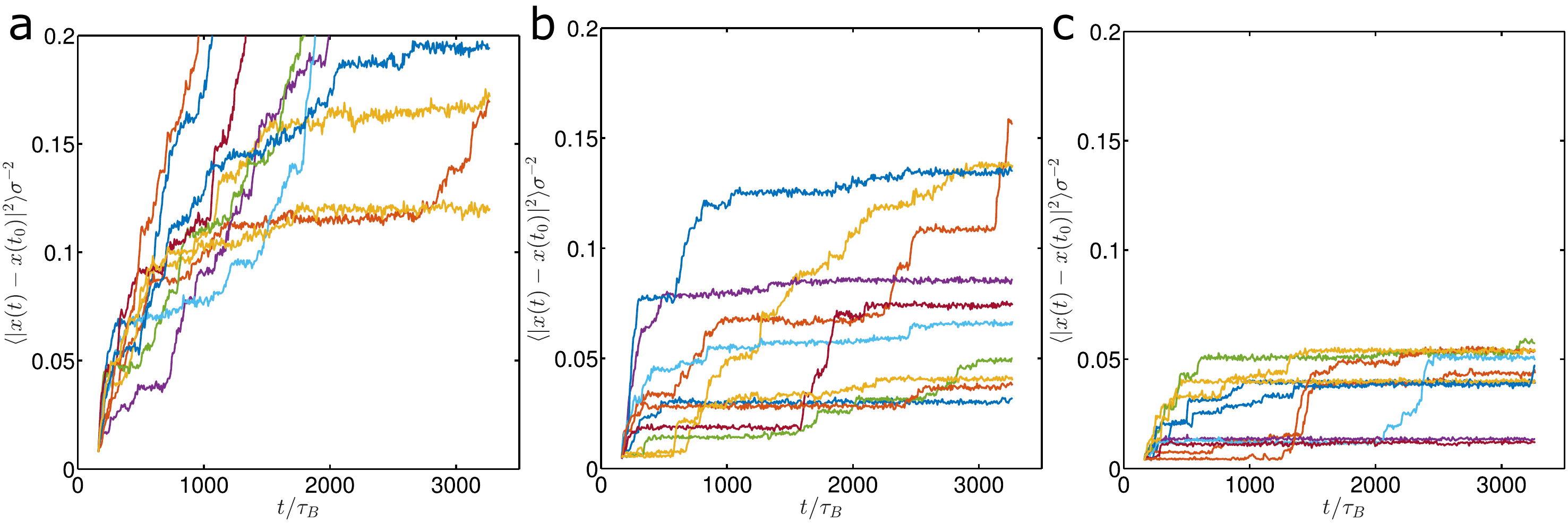}%
\end{center}
\caption{{\bf Likelihood of avalanche events vs. $\phi$.}
The mean-squared displacement of particles over time at different volume fractions for 10 independent trajectories. {\bf a,} 63\%; {\bf b,} 65\%; {\bf c,} 66\%. Different colours indicate different trajectories. There is a clear reduction in mobility, as well as an increased likelihood for the trajectories to stay in a metastable (plateau) state for more of the time domain shown.
}
\label{fig:MSDex}
\end{figure}

\begin{figure}[h]%
\begin{center}
\includegraphics[width=0.9\columnwidth]{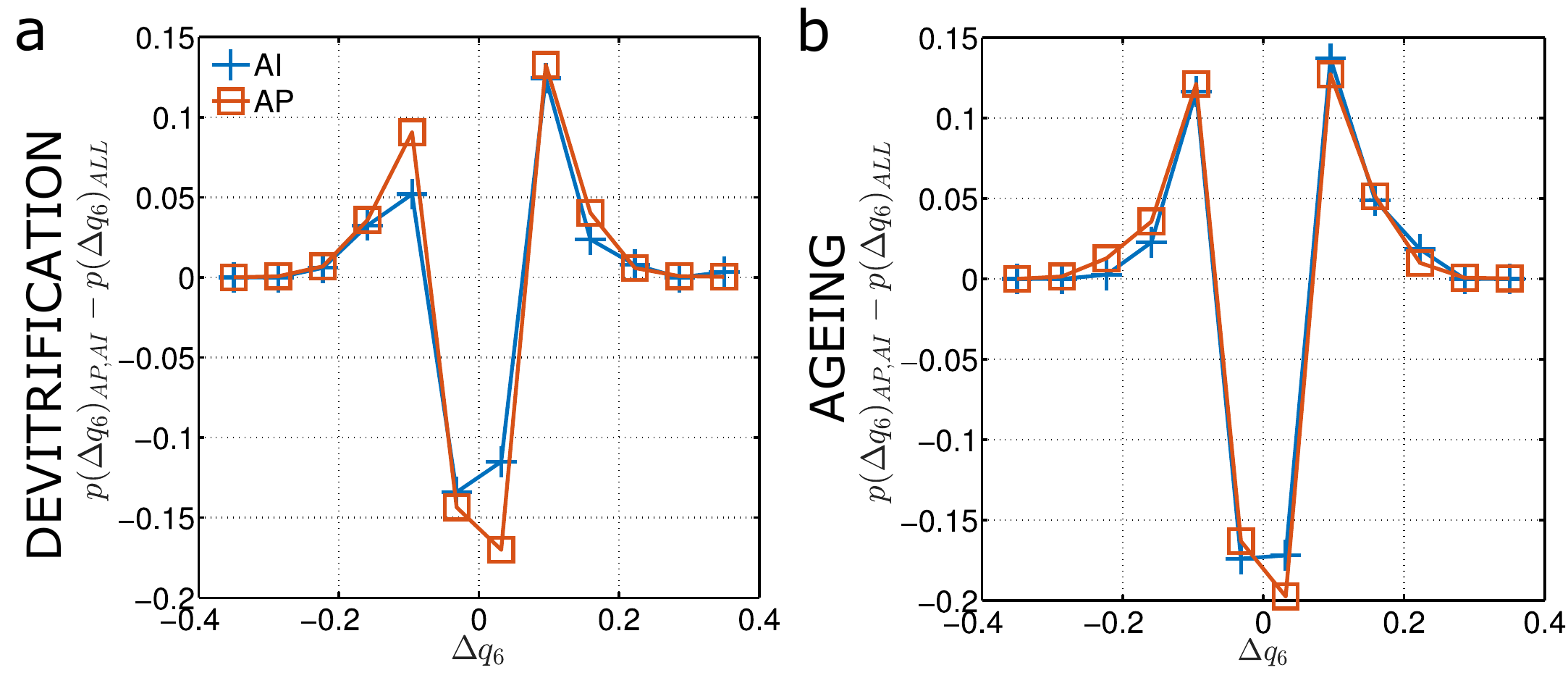}%
\end{center}
\caption{{\bf Change in $q_6$ for AIs and APs vs. all particles.}
{\bf a, b,} The difference between the probability of $\Delta{q_6}$ for AI/AP particles and the probability of $\Delta{q_6}$ for the entire particle population for devitrification and ageing events, respectively. Though the distinction between AIs and APs is clear for the monodisperse system, this is not the case for the polydisperse system (see the text).
}%
\label{fig:dq6}%
\end{figure}

\begin{figure}[h]%
\begin{center}
\includegraphics[width=0.6\columnwidth]{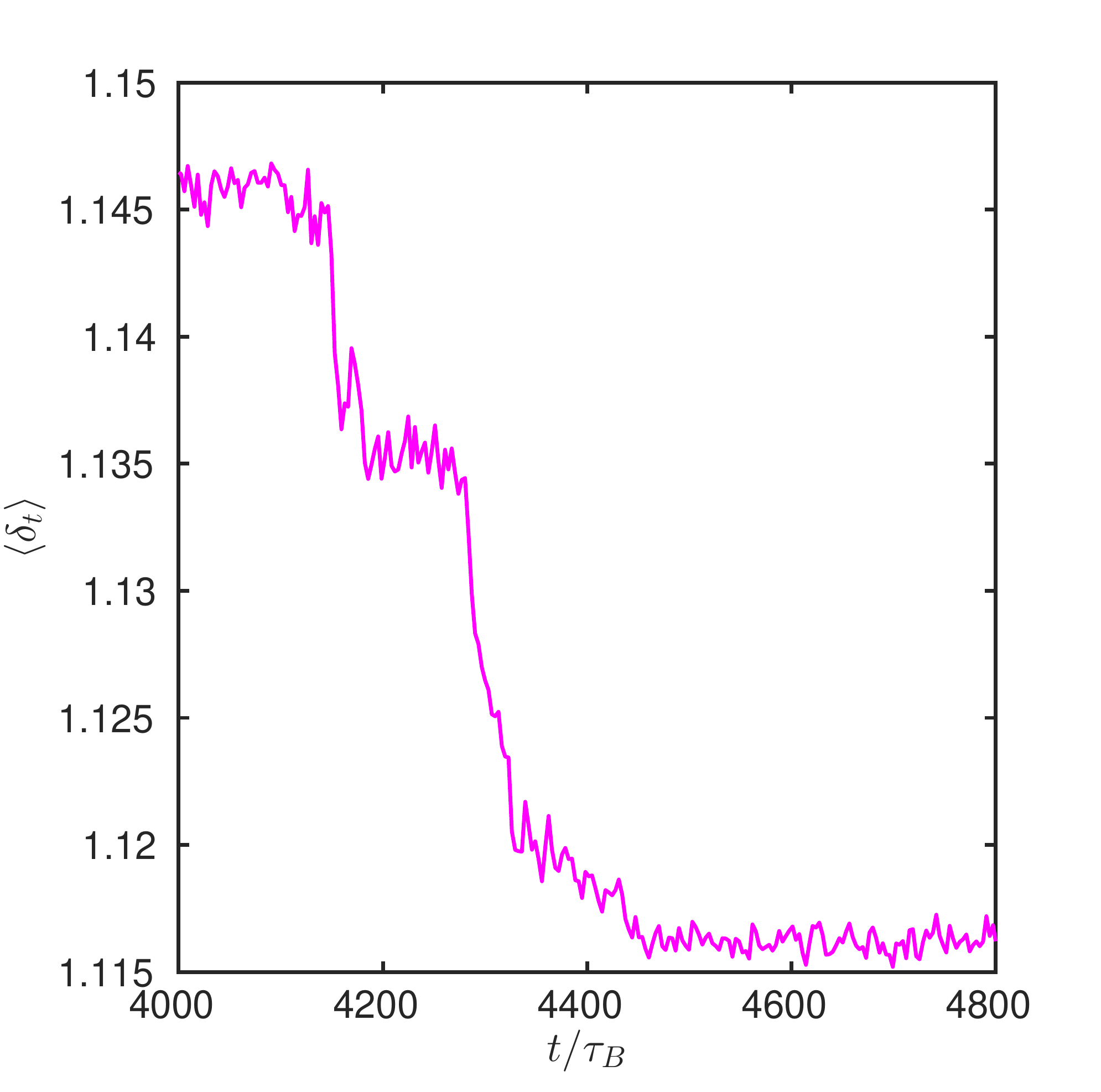}%
\end{center}
\caption{{\bf Collective motion.}
The average of the collective motion parameter $\delta$ over time, which decreases over the course of the avalanche. The event shown follows the same polydisperse trajectory as the one used for analysis of nearest neighbour and force neighbour changes in Fig.4a.
}%
\label{fig:Str}%
\end{figure}

\begin{figure}[h]%
\centering
\includegraphics[width=0.9\columnwidth]{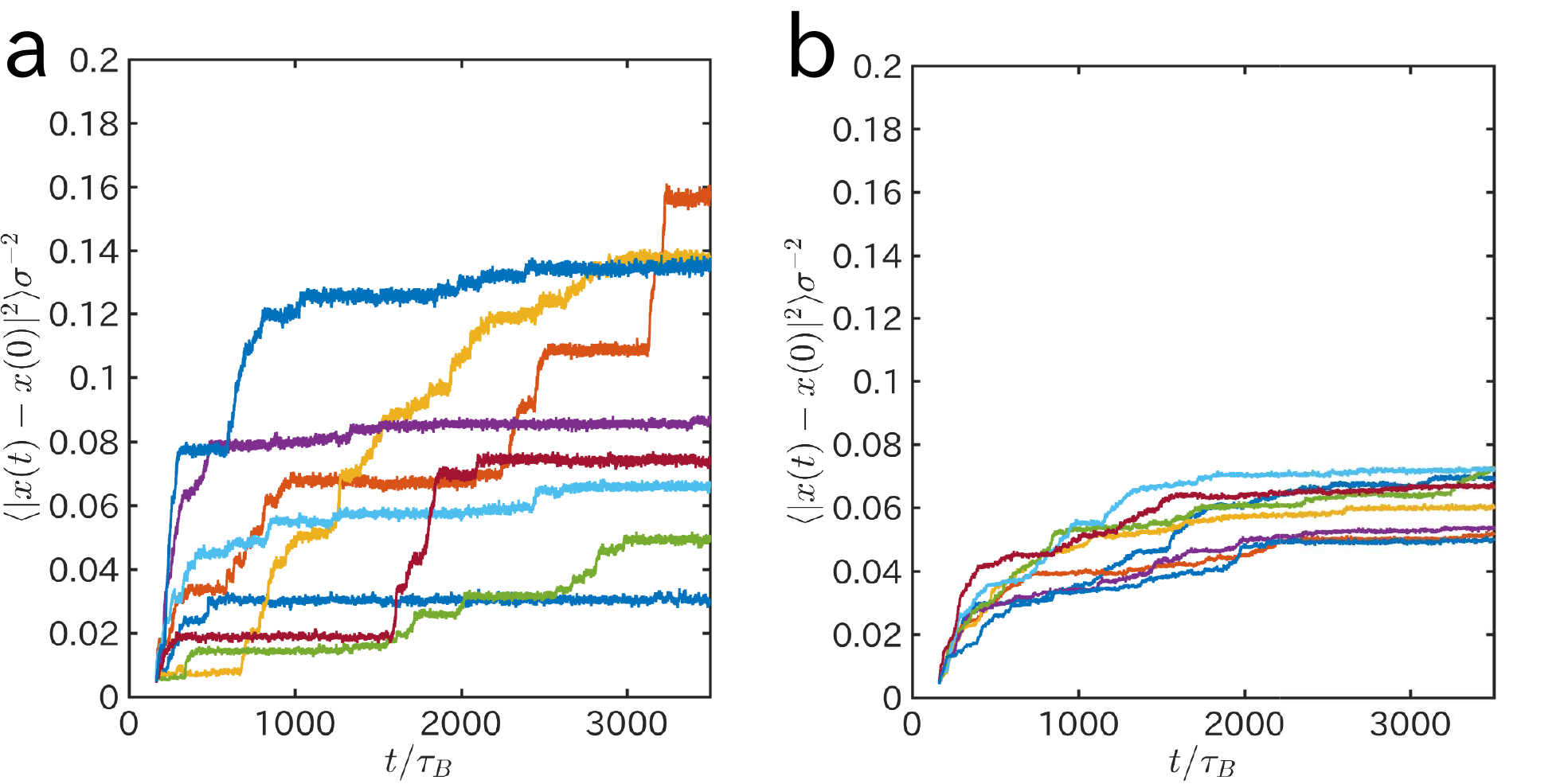}%
\caption{{\bf Mean-squared displacements for $n = 4000$ and $n = 16000$.}
MSDs found from eight independent trajectories for ({\bf a}) $n = 4000$ and ({\bf b}) $n = 16000$ system sizes. It can be seen that a larger system gives smaller events with less well-defined plateaux. This is due to events involving the same number of particles giving rise to smaller steps for larger $n$, and an increased chance of more events in a given time; events initiate while others have not relaxed yet.
}%
\label{fig:syssize}%
\end{figure}

\end{document}